\documentclass[english]{article}
\usepackage[T1]{fontenc}
\usepackage[latin9]{inputenc}
\usepackage[letterpaper]{geometry}
\usepackage{array}
\usepackage{textcomp}
\usepackage{graphicx}
\usepackage{fix-cm}
\makeatletter

\@ifundefined{definecolor}
 {\usepackage{color}}{}
\makeatother

\makeatother
\usepackage[T1]{fontenc}
\usepackage[latin9]{inputenc}
\usepackage{color}
\usepackage{textcomp}
\usepackage{amsmath}
\usepackage{amssymb}
\usepackage{graphicx}
\makeatother
\usepackage{babel}
\begin{document}

\title{Maximal entanglement concentration for $(n+1)$-qubit states}

\author{Anindita Banerjee $^{a}$ \footnote{email: aninditabanerjee.physics@gmail.com} \and Chitra Shukla $^{b}$   \and Anirban Pathak $^{b}$}

\maketitle
\begin{center}
$^{a}$Department of Physics and Center for Astroparticle Physics
and Space Science,
\par\end{center}
\begin{center}
Bose Institute, Block EN, Sector V, Kolkata 700091, India.
\par\end{center}
\begin{center}
$^{b}$Department of Physics and Materials Science and Engineering,
\par\end{center}
\begin{center}
Jaypee Institute of Information Technology, A-10, Sector-62, Noida,
UP-201307, India.
\par\end{center}
\maketitle




%
%
%
%

\begin{abstract}
We propose two  schemes for concentration of $(n+1)$-qubit entangled states that can be written in the form of $(\alpha|\varphi_{0}\rangle|0\rangle+\beta|\varphi_{1}\rangle|1\rangle)_{n+1}$
where $|\varphi_{0}\rangle$ and $|\varphi_{1}\rangle$ are mutually
orthogonal $n$-qubit states. The importance of this general form
is that the entangled states like Bell,  cat, GHZ, GHZ-like,
$|\Omega\rangle$, $|Q_{5}\rangle$, 4-qubit cluster states and specific
states from the 9 SLOCC-nonequivalent families of 4-qubit entangled
states can be expressed in this form. The
proposed entanglement concentration protocol is based on the local operations
and classical communications (LOCC). It is shown that the
maximum success probability for ECP using quantum nondemolition (QND)
technique is $2\beta^{2}$ for $(n+1)$-qubit states
of the prescribed form.  It is  shown that the proposed schemes can be implemented optically. Further it is also noted that the proposed schemes can be implemented using quantum dot and microcavity systems.



\end{abstract}

\section{Introduction\label{sec:Introduction}}

Entanglement is one of the most important resources of quantum information
processing. It plays a key role in quantum computation and communication. Specifically,
maximally entangled states are often a prerequisite for quantum information
processing. As entanglement between two or more distant parties cannot be created by local operations
and classical communication (LOCC), it is required to
be distributed. Usually a maximally entangled pure state is distributed
among different parties by quantum channels and due to the presence of noise it transforms to either a mixed state or a less entangled pure state.
Subsequently, maximally entangled state (MES) is extracted from an ensemble of mixed states or an ensemble of less entangled pure states. Entanglement
concentration is a process in which one can extract maximally entangled
state from nonmaximally (less entangled state) using LOCC with fidelity
equal to 1. Similarly, the process of distilling a set of mixed states into a maximally
entangled state is referred to as the  entanglement purification.

Entanglement concentration protocol (ECP) was proposed by Bennett
et al. \cite{origional ECP} in 1996. This pioneering protocol was based on the Schmidt
projection  method. Several ECPs were proposed thereafter using different
methods. For example, entanglement-swapping  \cite{S. Bose},
qubit-assisted \cite{bandyopadhyay}, POVM \cite{Bell-determinstic-ckt}
and Bell-measurement \cite{bellmeasurement} based ECPs were proposed. Initial  proposals for ECPs were restricted to the concentration of nonmaximally entangled Bell states.
However, with the recent advances in applications of multipartite entangled states in quantum communication
 and quantum computation, it has become extremely interesting to design ECPs for multipartite entangled states. For example, in the recent past,  ECPs
for various multipartite entangled states like GHZ and cat state \cite{Dhara GHZ state},
GHZ-like state \cite{chitraECP}, cluster state \cite{Dhara Cluster state,clusterCPL,clusterQIP},
arbitrary W state \cite{Sheng-2}, 4-qubit entangled state \cite{chitraECP},
etc. have been proposed.  Nevertheless, apart from using different
methods for designing ECPs, the possibility of implementation of ECPs is also studied for different
technologies, like linear optics \cite{Sheng-1,experiment2-zhao,experiment1-yamamoto,sheng_rev,renbc}, quantum dot and microcavity systems \cite{one,two,three,four}
and electronic technologies \cite{five,six}. Yamamoto et al. \cite{experiment1-yamamoto}
experimentally realized ECP from two copies of the state $\left|\varPhi\right\rangle =\alpha|00\rangle+\beta|11\rangle,$  where $|\alpha|^{2}+|\beta|^{2}=1$
using polarising beam splitters (PBSs), wave plates and photon detectors. The maximum probability
achieved in these cases for single pair concentration is $2\alpha^{2}\beta^{2}$.
Another system which is used in ECP  to obtain higher success probability
is the cross-Kerr-nonlinearity \cite{Sheng-2,Sheng-1}.

 A qubit-assisted ECP for Bell state
was initially proposed in \cite{bandyopadhyay} using CNOT gate and
Von Neumann measurement. There, it was  shown that using iterative process
maximal success probability probability up to $2\beta^{2}$ can be obtained for Bell state.
Recently, Sheng et al. \cite{Sheng-1} have proposed a qubit-assisted
concentration protocol for Bell state using PBS. They have shown that it is possible to obtain a higher success probability by iteratively
using cross-Kerr-nonlinearity.  Various qubit-assisted entanglement concentration
protocols \cite{q_a_chuang,q_a_cao,q_a_lan,GHZ-zhou,NOON,q_a_ybsheng,q_a_noon,q_a_hea}  for different entangled states using different systems have
been reported thereafter, referring to the higher success probability
mentioned in \cite{Sheng-1}. However, upper bound on the success probability was not discussed in these works. In what follows, we propose two  ECPs  for all
$(n+1)$-qubit entangled states that can be written in the form

\begin{equation}
\left(\alpha|\varphi_{0}\rangle|0\rangle+\beta|\varphi_{1}\rangle|1\rangle\right)_{n+1},\label{eq:general form}
\end{equation}
where $|\varphi_{0}\rangle$ and $|\varphi_{1}\rangle$ are mutually
orthogonal $n$-qubit states and $\alpha^{2}+\beta^{2}=1 : \alpha\neq\frac{1}{\sqrt{2}}$. Both the ECPs proposed here may be viewed as qubit-assisted ECPs as each of them is assisted by an ancillary qubit having  the same Schmidt coefficients as that of the less entangled state to be concentrated. Further, we show that one of the proposed ECP which utilizes QND techniques achieves the maximum possible success probability (i.e., $2\beta^{2}$). To justify the importance of the states of the form (\ref{eq:general form}), in Table \ref{tab:Entangled-states},
we have listed different entangled states which can be written in
the general form described in Eq. (\ref{eq:general form}).
Further, it may be noted that  the states of this particular form have
vast applications in secure quantum communication, quantum secret sharing
(QSS), bidirectional quantum teleportation and hierarchical quantum
communication schemes  \cite{HQIS,HQIS_mishra} e.g., hierarchical quantum information splitting
(HQIS), probabilistic HQIS and hierarchical quantum secret sharing
(HQSS).  The general nature and applicability of quantum states of
the form $\left(\alpha|\varphi_{0}\rangle|0\rangle+\beta|\varphi_{1}\rangle|1\rangle\right)_{n+1}$
have motivated us to construct ECPs for states of this form. Recently, we \cite{chitraECP} have proposed    a Bell state-assisted  ECP for the states of this particular form. Here we aim to propose two more ECPs for the same sates and to improve the efficiency.  The qubit-assisted ECP provides higher efficiency  as shown in  \cite{chitraECP} where efficiency of Bell state-assisted are compared with  qubit-assisted ECP. Moreover, the latter   requires  less resource compared to the  former. Further, if we  use nonlinear resources (i.e., if we use QND technique or equivalently if we use cross-Kerr nonlinearity) then the efficiency approaches allowed upper bound.

\begin{table}
\centering
\caption{Entangled states and their corresponding less entangled counter parts expressed in the general form described by Eq.
(\ref{eq:general form}).}
\label{tab:Entangled-states}
\begin{tabular}{lll}
\hline\noalign{\smallskip}
$(n+1)$-qubit state  & Maximally entangled state  & Non-maximally entangled state\\
\noalign{\smallskip}\hline\noalign{\smallskip}
2-qubit Bell  & $\frac{|00\rangle+|11\rangle}{\sqrt{2}}$  & $\left(\alpha|0\rangle|0\rangle+\beta|1\rangle|1\rangle\right)_{(1+1)}$\\

$(2+1)$-qubit GHZ  & $\frac{|000\rangle+|111\rangle}{\sqrt{2}}$  & $\left(\alpha|00\rangle|0\rangle+\beta|11\rangle|1\rangle\right)_{(2+1)}$\\

$(n+1)$-qubit cat  & $\frac{|0...00\rangle+|1...11\rangle}{\sqrt{2}}$  & $\left(\alpha|0...0\rangle|0\rangle+\beta|1...1\rangle|1\rangle\right)_{(n+1)}$\\

$(2+1)$-qubit~GHZ-like  & $\frac{|\psi^{+}0\rangle+|\phi^{+}1\rangle}{\sqrt{2}}$  & $\left(\alpha|\psi^{+}\rangle\left|0\right\rangle +\beta|\phi^{+}\rangle\left|1\right\rangle \right)_{(2+1)}$\\

$Q_{5}$ state  & $\frac{|0000\rangle+|1011\rangle+|1101\rangle+|1110\rangle}{2}$  & $(\alpha(|000\rangle+|111\rangle)|0\rangle$\\
  &  & $+\beta(|101\rangle+|110\rangle)|1\rangle)_{(3+1)}$\\

Cluster state  & $\frac{|0000\rangle+|0011\rangle+|1100\rangle-|1111\rangle}{2}$  & $(\alpha(|000\rangle+|110\rangle)|0\rangle$\\
  &   & $+\beta(|001\rangle-|111\rangle)|1\rangle)_{(3+1)}$\\
\noalign{\smallskip}\hline
\end{tabular}

\end{table}
Rest of the paper is organized as follows: In Section \ref{sec:Entanglement-concentration-protocols}, we have presented
our scheme.  In Section \ref{sec:Conservation-of-entanglement}, we have
discussed that the entanglement is conserved during ECP and finally
the paper is concluded in Section \ref{sec:Conclusion}.

\section{Maximal entanglement concentration \label{sec:Entanglement-concentration-protocols}}

We have proposed the ECP for $(n+1)$-qubit state
of the form
\[
\begin{array}{lcl}
 |\psi\rangle=\left(\alpha|\varphi_{0}\rangle|0\rangle+\beta|\varphi_{1}\rangle|1\rangle\right)_{n+1}.
\end{array}
\]

\begin{figure}
\begin{centering}
\includegraphics[scale=.7]{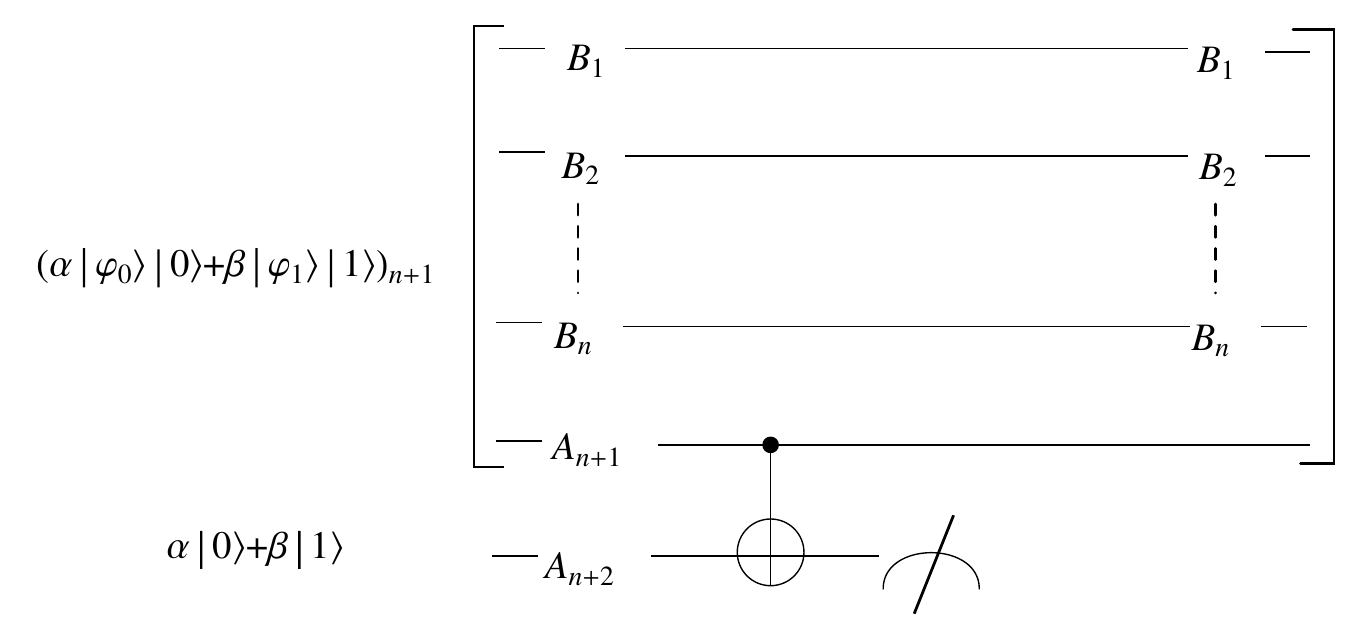}
\par\end{centering}

\protect\protect\caption{Proposed qubit-assisted maximal ECP for (n+1)-qubit state $\alpha|\psi_{0}\rangle|0\rangle+\beta|\psi_{1}\rangle|1\rangle$.\label{fig:Proposed-quantumcircuit}}
\end{figure}

The protocol can be implemented on all states shown in Table \ref{tab:Entangled-states}.
The protocol is described in Subsection \ref{sub:Quantum-circuit}
and the corresponding quantum circuit is shown in Fig. \ref{fig:Proposed-quantumcircuit}.
The purpose of presenting the quantum circuit is that the protocol
can be realized in any technology. We propose concentration
protocols using two independent methods: (i) post selection principle using
linear optics and (ii) quantum nondemolition using cross-Kerr-nonlinearity
as described in Subsection \ref{sub:Linear-Optics} and \ref{sub:Cross-Kerr-Nonlinearity},
respectively. Many concentration protocols have been realized using
linear optics as already stated in Section \ref{sec:Introduction}.
Linear optics has two important features first is post selection and
second is requirement of sophisticated single photon detectors. Post
selection destroys the photons  and they  cannot be further
used. The cross-Kerr-nonlinearity is widely studied in the context of the CNOT gate
\cite{cnot}, Bell state analysis \cite{bell state}, etc. It is a
tool to construct  quantum nondemolition detectors
(QND) having the capability  for conditioning the evolution of the
system without necessarily destroying the state of the photon. It acts as
parity check and single photon detector. Moreover, it is worthwhile
to mention that the protocol  here is implemented by linear optics,
but can be used in other technologies as well for example in electronic
systems like electronic polarization beam splitter and the charge
detection \cite{clusterQIP}, quantum dot spins \cite{chuang}, etc.
The quantum nondemolition technique implemented using these electronic
technologies have higher efficiency but the optimal value of success
probability has not been attained in their works. In this work we
show that
by employing QND technique we can attain an optimal value concentration
which is showed in Section \ref{sub:Cross-Kerr-Nonlinearity}.

\subsection{Quantum circuit\label{sub:Quantum-circuit}}

Let Alice and Bob be two spatially separated parties. They share $N$
number of $n+1$ pure entangled states such that Bob possesses first
$n$ qubits and Alice possesses $\left(n+1\right)^{th}$-qubit. The
entangled state is of the form $|\psi\rangle=\left(\alpha|\varphi_{0}\rangle|0\rangle+\beta|\varphi_{1}\rangle|1\rangle\right)_{n+1}$
where $|\varphi_{0}\rangle$ and $|\varphi_{1}\rangle$ are mutually
orthogonal $n$-qubit states. The values of  $\alpha$ and $\beta$ are real
such that $\alpha^{2}+\beta^{2}=1$ and $\alpha\neq\beta.$
Alice also possess one more qubit $|\psi_{s}\rangle=\alpha|0\rangle+\beta|1\rangle$ which has the same Schmidt coefficients
as that of the non-maximally entangled state shared by Alice and Bob.
Therefore, Alice possesses two qubits which are $(n+1)^{th}$
qubit and $(n+2)^{th}$ qubit.

Input state of the circuit  shown in Fig. \ref{fig:Proposed-quantumcircuit}  is
\begin{equation}
\begin{array}{lcl}
|\psi_{1}\rangle & = & |\psi\rangle\otimes|\psi_{s}\rangle\\
 & = & \left(\alpha|\varphi_{0}\rangle|0\rangle+\beta|\varphi_{1}\rangle|1\rangle\right)_{1,2,\cdots,n+1}\otimes(\alpha|0\rangle+\beta|1\rangle)_{n+2}\\
 & = & \left(\alpha^{2}|\varphi_{0}\rangle|00\rangle+\alpha\beta|\varphi_{1}\rangle|10\rangle\right.\\
 & + & \left.\alpha\beta|\varphi_{0}\rangle|01\rangle+\beta^{2}|\varphi_{1}\rangle|11\rangle\right)_{1,2,\cdots,n+1,n+2}.
\end{array}\label{eq:qc1}
\end{equation}
Alice applies a CNOT gate on her qubits by using $(n+1)^{th}$ qubit
as the control qubit and the $(n+2)^{th}$ qubit as the target qubit.
As a consequence of this operation input state $|\psi_{1}\rangle$
transforms to
\begin{equation}
\begin{array}{lcl}
\begin{array}{c}
|\psi_{2}\rangle\end{array} & = & \left(\alpha^{2}|\varphi_{0}\rangle|00\rangle+\alpha\beta|\varphi_{1}\rangle|11\rangle\right.\\
 & + & \left.\alpha\beta|\varphi_{0}\rangle|01\rangle+\beta^{2}|\varphi_{1}\rangle|10\rangle\right)_{1,2,\cdots,n+1,n+2}\\
 & = & \left[\left(\alpha^{2}|\varphi_{0}\rangle|0\rangle+\beta^{2}|\varphi_{1}\rangle|1\rangle\right)|0\rangle\right.\\
 & + & \left.\alpha\beta\left(|\varphi_{0}\rangle|0\rangle+|\varphi_{1}\rangle|1\rangle\right)|1\rangle\right]_{1,2,\cdots,n+1,n+2}.
\end{array}\label{eq:qc2}
\end{equation}
Alice measures $(n+2)^{th}$ qubit of $|\psi_{2}\rangle$ in the computational
$\left\{ |0\rangle,|1\rangle\right\} $ basis. From (\ref{eq:qc2}),
we can easily observe that if Alice's measurement yields $|1\rangle$
then the quantum state shared by Alice and Bob collapses to a normalized
maximally entangled state $\frac{|\varphi_{0}\rangle|0\rangle+|\varphi_{1}\rangle|1\rangle}{\sqrt{2}}$,
but the circuit (equivalently an ECP represented by the circuit) fails
when Alice's measurement yields $|0\rangle$. Clearly, the success
probability of the ECP is $2\alpha^{2}\beta^{2}$ \textcolor{black}and
we are now left with a less entangled pure state with probability
$1-2\alpha^{2}\beta^{2}$  i.e., $\alpha^{4}+\beta^{4}$. We can repeat
the ECP using the remaining less entangled pure states to maximize
 the number of maximally
entangled states. The maximum
probability that can be achieved by repeating this procedure is $\mbox{2\ensuremath{\beta^{2}}}$ which
is provided later in Subsection \ref{sub:Cross-Kerr-Nonlinearity}.

\subsection{Entanglement concentration using linear optics\label{sub:Linear-Optics}}

\begin{figure}
\centering{}\includegraphics[scale=.55]{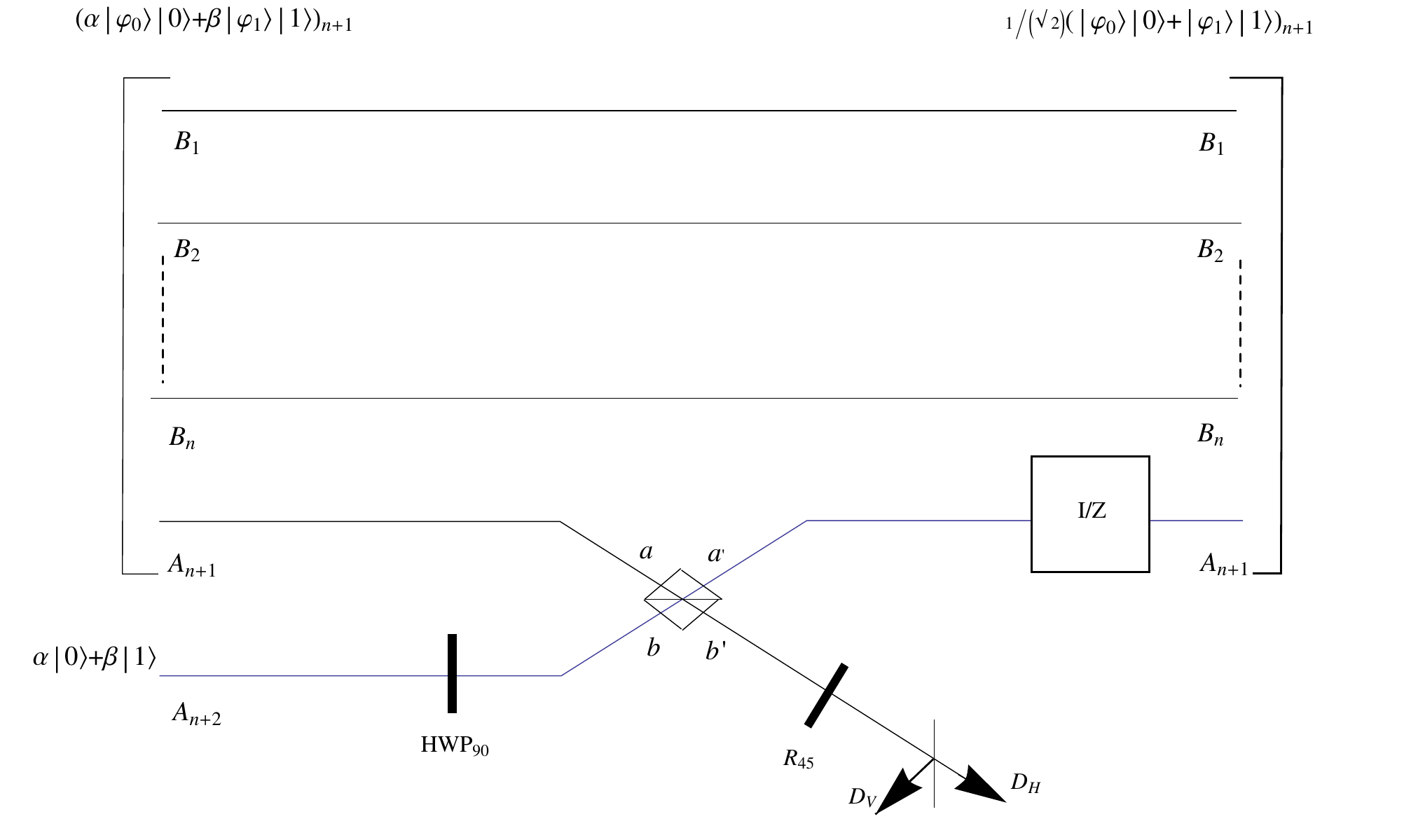}\caption{\label{fig:Proposed-linear-optics}Proposed linear optics circuit
for ECP for non-maximally entangled $(n+1)$-qubit state $\alpha|\varphi_{0}\rangle|0\rangle+\beta|\varphi_{1}\rangle|1\rangle$.}
\end{figure}

In optical implementation of an ECP, we consider that qubits are realized
using polarization states of the photon where horizontal ($H$) and
vertical ($V$) photon represent the logical bits $0$ and $1.$ Thus,
the quantum state to be concentrated can be expressed as $\alpha|\varphi_{0}H_{a}\rangle+\beta|\varphi_{1}V_{a}\rangle.$
A linear optics based scheme for entanglement concentration of this
state is shown in Fig. \ref{fig:Proposed-linear-optics} and the
action of the same is described below. Initial state of the system
can be expressed as

\begin{equation}
\begin{array}{lcl}
|\psi_{{\rm 3}}\rangle=|\psi\rangle\otimes|\psi_{s}\rangle & = & \left[\alpha|\varphi_{0}H_{a}\rangle+\beta|\varphi_{1}V_{a}\rangle\right.\\
 &  & \left.\otimes(\alpha|H_{b}\rangle+\beta|V_{b}\rangle)\right]_{n+2}\\
 & = & \alpha^{2}|\varphi_{0}H_{a}H_{b}\rangle+\beta^{2}|\varphi_{1}V_{a}V_{b}\rangle\\
 &  & +\alpha\beta|\varphi_{0}H_{a}V_{b}\rangle+\alpha\beta|\varphi_{1}V_{a}H_{b}\rangle.
\end{array}\label{eq1}
\end{equation}
Here, subscripts $a$ and $b$ represent the modes of  Alice's photon.
To be precise, $a$ corresponds to the $(n+1)^{th}$ qubit of the
shared entangled state and $b$ corresponds to the additional single
photon state ($(n+2)^{th}$ qubit) prepared by Alice to implement
the ECP. Now, Alice uses a half wave plate ${\rm HWP_{90}}$ to rotate
the polarization of photon in mode $b$ by $90$ degrees to transform
$|\psi_{{\rm 3}}\rangle$ to

\begin{equation}
\begin{array}{cc}
\begin{array}{ccc}
|\psi_{{\rm 4}}\rangle & = & \alpha^{2}|\varphi_{0}H_{a}V_{b}\rangle+\beta^{2}|\varphi_{1}V_{a}H_{b}\rangle+\\
 &  & \alpha\beta|\varphi_{0}H_{a}H_{b}\rangle+\alpha\beta|\varphi_{1}V_{a}V_{b}\rangle.
\end{array}\end{array}\label{eq:2}
\end{equation}
Photons in mode $a$ and $b$ enter polarized beam splitter (PBS).
As the PBS transmits the horizontal polarization component and reflects
the vertical polarization component after the PBS the state of the
system transforms to

\begin{equation}
\begin{array}{ccc}
|\psi_{5}\rangle & = & \alpha^{2}|\varphi_{0}H_{b^{\prime}}V_{b^{\prime}}\rangle+\beta^{2}|\varphi_{1}V_{ a^{\prime}}H_{ a^{\prime}}\rangle\\
 &  & \begin{array}{cc}
 & +\alpha\beta|\varphi_{0}H_{b^{\prime}}H_{a^{\prime}}\rangle+\alpha\beta|\varphi_{1}V_{a^{\prime}}V_{b^{\prime}}\rangle.\end{array}
\end{array}\label{eq:3}
\end{equation}

Note that in the first two terms in (\ref{eq:3}), the last two photons
are in the same spatial mode, whereas in the last two terms one photon
is present in each output mode. Alice selects the case where each
mode contains one photon. Therefore, after this choice Alice and Bob
obtain a shared state $\frac{1}{\sqrt{2}}\left(|\varphi_{0}\rangle|H_{a^{\prime}}H_{b^{\prime}}\rangle+|\varphi_{1}\rangle|V_{a^{\prime}}V_{b^{\prime}}\rangle\right)_{n+2}$
with probability $2\alpha^{2}\beta^{2}$. Now, Alice allows the photon
in mode $b'$ to pass through $R_{45}$ which rotates the polarization
by $45$ degrees. As a consequence the state of the system transforms
to
\[
\begin{array}{lcl}
|\psi_{6}\rangle & = & \frac{1}{2}\left(|\varphi_{0}\rangle|H_{a^{\prime}}H_{b^{\prime}}\rangle+|\varphi_{0}\rangle|H_{a'a^{\prime}}V_{b^{\prime}}\rangle\right.\\
 & + & \left.|\varphi_{1}\rangle|V_{a^{\prime}}H_{b^{\prime}}\rangle-|\varphi_{1}\rangle|V_{a^{\prime}}V_{b^{\prime}}\rangle\right)_{n+2}\\
 & = & \frac{1}{2}\left(\left(|\varphi_{0}\rangle|H_{a^{\prime}}\rangle+|\varphi_{1}\rangle|V_{a^{\prime}}\rangle\right)\left|H_{b^{\prime}}\right\rangle \right.\\
 & + & \left.\left(|\varphi_{0}\rangle|H_{a^{\prime}}\rangle-|\varphi_{1}\rangle|V_{a^{\prime}}\rangle\right)\left|V_{b^{\prime}}\right\rangle \right)_{n+2}.
\end{array}
\]
Finally, if the detector detects $(n+2)^{th}$ photon in the horizontal
mode (i.e., if $D_{H}$ clicks) then Alice and Bob obtain the state

\[
\begin{array}{lcl}
 \frac{1}{\sqrt{2}}\left(|\varphi_{0}\rangle|H\rangle+|\varphi_{1}\rangle|V\rangle\right)_{n+1}
 \end{array}
 \]
and similarly if the detector $D_{V}$ clicks, then they obtain the
state
\[
\begin{array}{lcl}
 \frac{1}{\sqrt{2}}\left(|\varphi_{0}\rangle|H\rangle-|\varphi_{1}\rangle|V\rangle\right)_{n+1}.
 \end{array}
 \]
It is easy to observe that on application of the unitary operation
in both the cases we obtain a maximally entangled state $\frac{1}{\sqrt{2}}\left(|\varphi_{0}\rangle|0\rangle+|\varphi_{1}\rangle|1\rangle\right)_{n+1}$.
Therefore, the probability $P_{1}$ to obtain the maximally entangled
state
$\frac{1}{\sqrt{2}}\left(|\varphi_{0}\rangle|H\rangle+|\varphi_{1}\rangle|V\rangle\right)_{n+1}$  or $\frac{1}{\sqrt{2}}\left(|\varphi_{0}\rangle|H\rangle-|\varphi_{1}\rangle|V\rangle\right)_{n+1}$ is $2(\alpha\beta)^{2}$.

\subsection{Entanglement concentration  using QND detectors (cross-Kerr-nonlinearity)\label{sub:Cross-Kerr-Nonlinearity}}
\begin{figure}
\centering{}\includegraphics[scale=1]{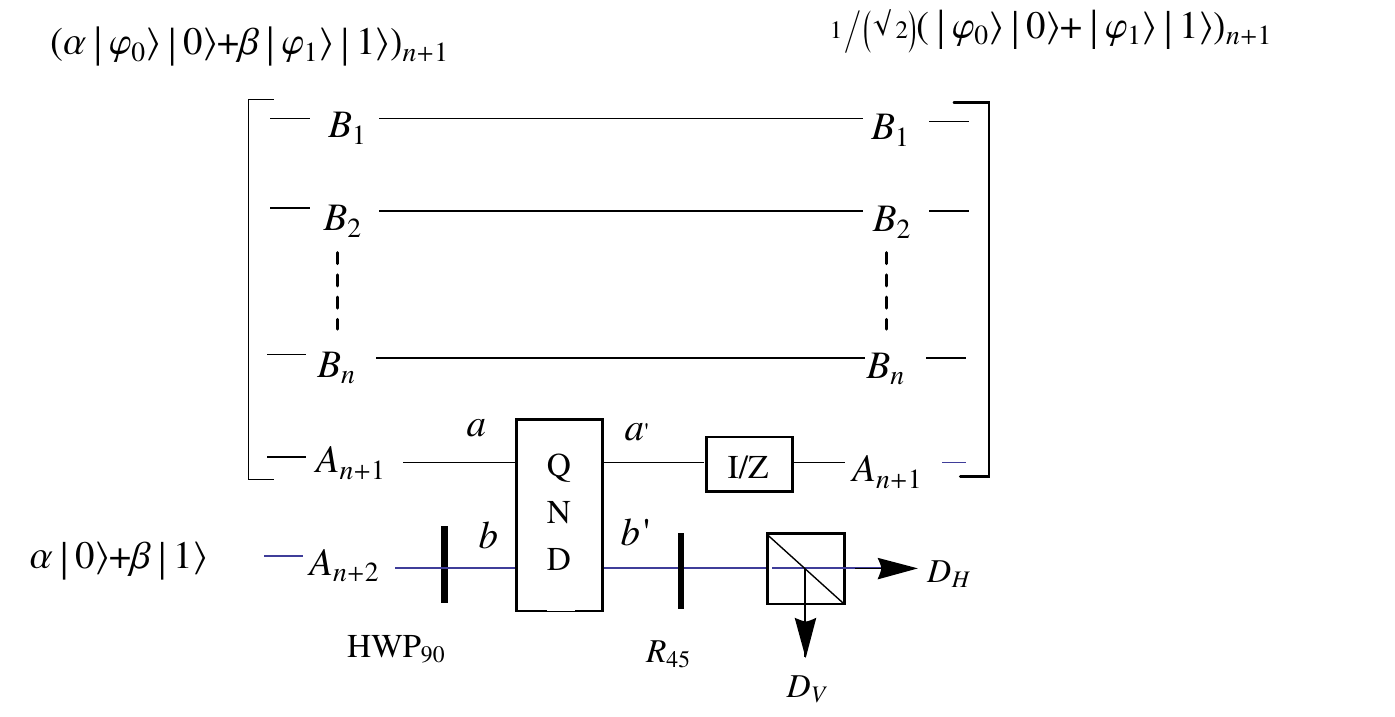}\protect\protect\caption{Schematic of ECP with QND.\label{fig:proposedQND}}
\end{figure}

Quantum nondemolition detector (QND) is generally
based on cross-Kerr-nonlinearity. After their introduction by Nemoto
and Munro \cite{cnot} many works precisely in entanglement concentration protocols of pure entangled states were reported. The Hamiltonian of the nonlinear medium or interaction between the
signal beam and probe beam can be written as
\begin{equation}
\hat{H}=\hbar\chi\hat{n}_{s}\hat{n}_{p}\label{eq:hamiltonian}
\end{equation} where $\hbar\chi$ is the coupling strength of the cross-Kerr material,
$\hat{n}_{s}$ and $\hat{n}_{p}$ are the number operators of signal
mode $s$ and probe mode $p$, respectively. Let
the signal mode be in input state $|\varphi\rangle=a|H\rangle+b|V\rangle$
and the probe mode be in a coherent state $|\alpha\rangle$. The cross-Kerr
nonlinear medium causes the combined input state of $|\varphi\rangle$
and coherent state $|\alpha\rangle$ to evolve as

\begin{equation}
\begin{array}{ccc}
U_{cross-Kerr}|\varphi\rangle|\alpha\rangle & = & e^{iHt/\hbar}(a|0\rangle+b|1\rangle)|\alpha\rangle\\
 & = & a|0\rangle|\alpha\rangle+b|1\rangle|\alpha e^{i\theta}\rangle,
\end{array}\label{eq:crosskerr}
\end{equation}
where $\theta=\chi t$ and $t$ is the interaction time. In Eq.
(\ref{eq:crosskerr}), $       |n\rangle :n\in{0,1}  $ represents the
Fock state  contains  $n$ photons.
\begin{figure}
\begin{centering}
\includegraphics[scale=.7]{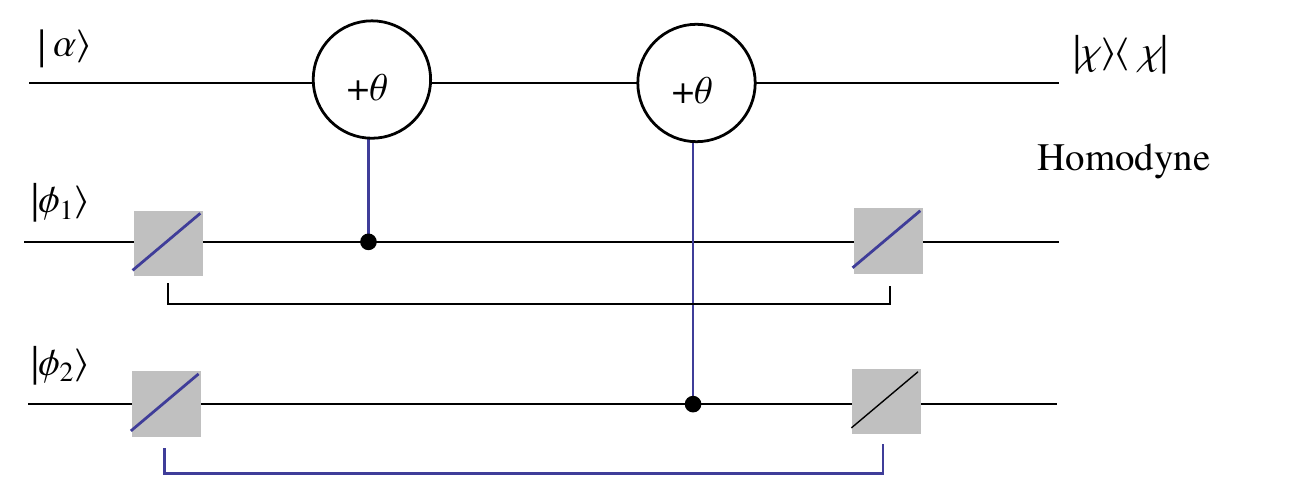}
\par\end{centering}

\caption{Schematic of the  quantum nondemolition detector (QND)
constructed by cross-Kerr-nonlinearity to distinguish between $|HH\rangle$
and $|VV\rangle$ from $|HV\rangle$ and $|VH\rangle$.\label{fig:SchematicQND}}
\end{figure}
Therefore, we observe that $|\varphi\rangle$ remains same and the probe
state picks up a phase shift and it can be measured by homodyne detection.
The phase shift is directly proportional to the number of photons.
In this way, we can nondestructively determinate  the state of the signal mode  by measuring the probe mode. Thus, QND can condition
the evolution of the system by not destroying the single photon. This
is non-destructive measurement.
QND measurement is used for parity checking i.e., to differentiate
between a pair of identical polarizations $|HH\rangle$ or $|VV\rangle$
from a pair of orthogonal polarizations $|HV\rangle$ or $|VH\rangle$
depending on the phase shift $\theta$ of the coherent state. QND
based scheme for parity checking is shown in Fig. \ref{fig:SchematicQND}.   This scheme is frequently used in quantum information and detail description of this scheme can be found in \cite{sheng_rev,cnot}.
Let $|\phi_{1}\rangle$ and $|\phi_{2}\rangle$ be two polarization
qubits.

\[
\begin{array}{c}
|\phi_{1}\rangle=a_{1}|H\rangle+b_{1}|V\rangle\\
|\phi_{2}\rangle=a_{2}|H\rangle+b_{2}|V\rangle
\end{array}
\]
The composite quantum system is $|\phi_{1}\rangle\otimes|\phi_{2}\rangle\otimes|\alpha\rangle$.
This system will evolve to
\[
\begin{array}{c}
\left[a_{1}a_{2}|HH\rangle+b_{1}b_{2}|VV\rangle\right]|\alpha e^{i\theta}\rangle+a_{1}b_{2}|HV\rangle|\alpha e^{i2\theta}\rangle\\
+b_{1}a_{2}|VH\rangle|\alpha\rangle
\end{array}
\]
The qubit is passed through PBS and enters the spatial mode which interacts
with nonlinear medium. The coherent beam picks up a phase shift $\theta$
if there is $|HH\rangle$ or $|VV\rangle$ and $2\theta$ or no phase
shift if there are $|HV\rangle$ or $|VH\rangle$ in a mode, respectively.
Our proposed scheme is shown in Fig. \ref{fig:proposedQND}, where
we place a QND measurement based parity checking box in place of the
first PBS used in the linear optics based scheme described earlier
(cf. Fig. \ref{fig:Proposed-linear-optics}).
The input state in Fig. \ref{fig:proposedQND} is
\[
\begin{array}{lcl}
|\psi_{{\rm 7}}\rangle & = & |\psi_{4}\rangle|\alpha\rangle\\
 & = & \left(\alpha^{2}|\varphi_{0}H_{a}V_{b}\rangle+\beta^{2}|\varphi_{1}V_{a}H_{b}\rangle\right.\\
 & + & \left.\alpha\beta|\varphi_{0}H_{a}H_{b}\rangle+\alpha\beta|\varphi_{1}V_{a}V_{b}\rangle\right)|\alpha\rangle.
\end{array}
\]
After operation of the QND measurement based parity checking box the
state transforms to
\[
\begin{array}{lcl}
|\psi_{{\rm 8}}\rangle & = & \alpha^{2}|\varphi_{0}H_{a}V_{b}\rangle|\alpha e^{i2\theta}\rangle+\beta^{2}|\varphi_{1}V_{a}H_{b}\rangle|\alpha\rangle\\
 & + & \alpha\beta(|\varphi_{0}H_{a}H_{b}\rangle+|\varphi_{1}V_{a}V_{b}\rangle)|\alpha e^{i\theta}\rangle.
\end{array}
\]
QND based scheme for ECP of the same state is shown in the Fig. \ref{fig:SchematicQND}. Therefore, if there is a phase shift of $\theta$ then Alice will
communicate to Bob to keep the state. After the measurement if $D_{H}$
clicks, then Alice and Bob obtain maximally entangled state $\frac{1}{\sqrt{2}}\left(|\varphi_{0}\rangle|H\rangle+|\varphi_{1}\rangle|V\rangle\right)_{n+1}$
as desired. Similarly, if $D_{V}$ clicks, then the combined state
of Alice and Bob, collapses to $\frac{1}{\sqrt{2}}\left(|\varphi_{0}\rangle|H\rangle-|\varphi_{1}\rangle|V\rangle\right)_{n+1}$
which can be transformed to the desired maximally entangled state
by a phase flip operation. Thus, if $D_{V}$ clicks, then Bob applies
a phase flip operation to obtain the desired state.
Now the states for which phase shift is either $2\theta$
or no phase shift form the less entangled states and are given by
\[
\alpha^{\prime}|\varphi_{0}H_{a^{\prime}}V_{b^{\prime}}\rangle+\beta^{\prime}|\varphi_{1}V_{a^{\prime}}H_{b^{\prime}}\rangle,
\]
where $\alpha^{\prime}=\frac{\alpha^{2}}{\sqrt{\alpha^{4}+\beta^{4}}}$
and $\beta^{\prime}=\frac{\beta^{2}}{\sqrt{\alpha^{4}+\beta^{4}}}$. The
qubit in mode $b'$ passes through $R_{45}$ and as a consequence,
we obtain
\[
\begin{array}{c}
\left(\alpha^{\prime}|\varphi_{0}H_{a^{\prime}}\rangle+\beta^{\prime}|\varphi_{1}V_{a^{\prime}}\rangle\right)|H_{b^{\prime}}\rangle-\left(\alpha^{\prime}|\varphi_{0}H_{a^{\prime}}\rangle-\beta^{\prime}|\varphi_{1}V_{a'}\rangle\right)|V_{b'}\rangle.
\end{array}
\]
Therefore, if detector $D_{H}$ clicks we get
the state $\alpha^{\prime}|\varphi_{0}H_{a^{\prime}}\rangle+\beta^{\prime}|\varphi_{1}V_{a^{\prime}}\rangle$
and Alice needs to prepare another single photon
having the same Schmidt coefficients i.e., $\alpha^{\prime}|H_{a2}\rangle+\beta^{\prime}|V_{a2}\rangle$.
Alice can reconstruct the maximally entangled state following the same
process. The probability of success of second iteration is given by
\[
(1-P_{1})P_{2}=(1-2\alpha^{2}\beta^{2})\frac{2(\alpha^{\prime}\beta^{\prime})^{2}}{\left((\alpha^{\prime})^{2}+(\beta^{\prime})^{2}\right)^{2}}=\frac{2\alpha^{4}\beta^{4}}{\alpha^{4}+\beta^{4}}.
\]
Thus, the probability of success after $N$ iterations would be

\[
P=P_{1}+(1-P_{1})\left[P_{2}\right.+(1-P_{2})\left[P_{3}\right.\cdots+\left.(1-P_{N-1})\left[P_{N}\right]\right].
\]
This can be written as
\[
\begin{array}{lcl}
P & = & 2\alpha^{2^{1}}\beta^{2^{1}}+\frac{2\alpha^{2^{2}}+\beta^{2^{2}}}{\left(\alpha^{2^{2}}+\beta^{2^{2}}\right)}+\frac{2\alpha^{2^{3}}+\beta^{2^{3}}}{\left(\alpha^{2^{2}}+\left|\beta\right|^{2^{2}}\right)\left(\alpha^{2^{3}}+\beta^{2^{3}}\right)}+\cdots\\
 & + & \frac{2\alpha^{2^{N}}\beta^{2^{N}}}{\left(\alpha^{2^{2}}+\beta^{2^{2}}\right)\left(\alpha^{2^{3}}+\beta^{2^{3}}\right)\left(\alpha^{2^{4}}+\beta^{2^{4}}\right)\cdots\left(\alpha^{2^{N}}+\beta^{2^{N}}\right)}\\
 & = & 2\alpha^{2}\beta^{2}+2\beta^{4}\left[\frac{x^{2^{2}-4}}{\left(1+x^{2^{2}}\right)}+\frac{x^{2^{3}-4}}{\left(1+x^{2^{2}}\right)\left(1+x^{2^{3}}\right)}+\right.\\
 & + & \left.\cdots\frac{x^{2^{N-4}}}{\left(1+x^{2^{2}}\right)\cdots\left(\left|\alpha\right|^{2^{N}}+\left|\beta\right|^{2^{N}}\right)}\right],
\end{array}
\]
where $x=\frac{\beta}{\alpha}$ and $0<x<1.$ For large $N,$ i.e.,
$N\rightarrow\infty,$ above series (the series present within square
bracket of the expression of $P$) becomes an infinite series and
that converges to $1$ for $x\in(0,1).$ Therefore, $P=2\beta^{2}\left(\alpha^{2}+\beta^{2}\right)=2\beta^{2}.$
Thus, the maximum probability with which we can obtain a MES from
the corresponding less entangled state for $(n+1)$-qubit state of
the form $|\psi\rangle=\left(\alpha|\varphi_{0}\rangle|0\rangle+\beta|\varphi_{1}\rangle|1\rangle\right)_{n+1}$
is $2\beta^{2}.$ It is a straightforward exercise to show that
the same bound on the maximum success probability is also applicable
to the recent proposals of Sheng et. al. \cite{Sheng-2,Sheng-1}.

\section{Conservation of entanglement\label{sec:Conservation-of-entanglement}}
It is well known that there exists a large number of measures of entanglement, each having its merit and demerit.
In the investigations related to entanglement concentration and purification, a specific measure known as \emph{entanglement of
single pair purification} (ESPP) is often used \cite{S. Bose,bandyopadhyay}. This measure was introduced by Bose et al. \cite{S. Bose} as
 the   maximum probability with which two parties sharing a non-maximally entangled pure Bell-type state
can  transform it to a Bell state by using LOCC (local operation and classical communication). Historically, Bose et al., defined ESPP
in context of Bell states as in their original work they described an ECP for Bell state and showed that ESPP is such a measure of
entanglement which remains conserved in the ECP described by them. Later on Bandyopadhyay \cite{bandyopadhyay} showed that ESPP remains conserved
in a qubit-assisted ECP for Bell state. Motivated by this observation, here we plan to show that ESPP is also conserved in the QND-based ECP proposed here. To do so, first we need to generalize the definition of ESPP for states
 other than Bell state as our ECP is valid for all states of the form described in Eq. (\ref{eq:general form}). The generalized definition of ESPP is as follows: ESPP is  the   maximum probability with which two parties sharing a non-maximally entangled pure entangled  state
can  transform it to the corresponding maximally entangled state by using LOCC.
Following Lo and Popescu \cite{Popescu} we may note that the
 maximum probability with which a nonmaximally entangled pure state of the form provided in Eq. (\ref{eq:general form}) can be concentrated
is twice the modulus square of the Schmidt coefficient of the smaller
magnitude. Therefore, before  concentration   the average entanglement
shared between Alice and Bob is $\left\langle E_{{\rm before}}\right\rangle _{AB}=$$2\beta^{2}$
where $\alpha>\beta$. In the proposed QND based protocol,
the maximum probability for $(n+1)$-qubit state after the concentration
is $\left\langle E_{{\rm after}}\right\rangle _{AB}=2\alpha^{2}\beta^{2}+2\beta^{4}$
$=2\beta^{2}$. This implies that the average ESPP is conserved, the maximum probability to
concentrate $(n+1)$-qubit state is $2\beta^{2}$ and that the remaining
will be totally disentangled if the protocol is continued for sufficiently long time.
Here we have followed the works of Bose et al. \cite{S. Bose}  and Bandyopadhyay  \cite{bandyopadhyay} to establish the conservation of entanglement
in the similar sense. However,
it may be noted that this conclusion (i.e., conservation of entanglement) is entanglement measure specific and it may
and may not hold for another measure of entanglement.

\section{Conclusion\label{sec:Conclusion}}

We have proposed two practical schemes for entanglement concentration
for all entangled states that can be expressed in the form  $\left(\alpha|\varphi_{0}\rangle|0\rangle+\beta|\varphi_{1}\rangle|1\rangle\right)_{n+1}$.
Specifically, we have proposed an ECP using linear optics i.e., using
PBSs and single photon detectors. Subsequently, we have also proposed
an ECP using cross-Kerr-nonlinearity and have obtained the maximum possible
success probability $2\beta^{2}$.  The proposed ECP
uses less resources compared to the earlier proposals.   We have
presented a quantum circuit for the proposed ECP so  that the protocol can be implemented using other technologies
like atomic system, NMR, etc. The proposed ECP can be applied to many entangled
states like Bell and cat states, GHZ, GHZ-like,
$|\Omega\rangle$, $|Q_{5}\rangle$, 4-qubit cluster states and specific
states from the 9 SLOCC-nonequivalent families of 4-qubit entangled
states and other states which can be expressed
as $\left(\alpha|\varphi_{0}\rangle|0\rangle+\beta|\varphi_{1}\rangle|1\rangle\right)_{n+1}$.

\section*{Acknowledgement}

AB thanks DST-SERB project SR/S2/LOP-18/2012. AB also acknowledges
S. Bandyopadhyay for some technical discussion and his interest in
the present work. AP thanks Department of Science and Technology (DST),
India for support provided through the DST project No. SR/S2/LOP-0012/2010

\end{document}